\documentclass{elsart}
\usepackage{epsfig}
\begin{document}
\runauthor{Cicero, Caesar and Vergil}
\begin{frontmatter}
\title{Isospin dependent properties of asymmetric nuclear matter}
\vskip -0.7cm

\author[SINP]{P. Roy Chowdhury\thanksref{X}},
\author[VECC]{D.N. Basu\thanksref{Z}} and
\author[SINP,VCU,UR]{C. Samanta\thanksref{Y}}

\vskip -0.25cm

\address[SINP]{Saha Institute of Nuclear Physics, 1/AF Bidhan Nagar, Kolkata 700 064, India }
\address[VECC]{Variable  Energy  Cyclotron  Centre, 1/AF Bidhan Nagar, Kolkata 700 064, India }
\address[VCU]{Physics Department, Virginia Commonwealth University, Richmond, VA 23284-2000, U.S.A.}
\address[UR]{Physics Department, University of Richmond, VA 23173, U.S.A.}

\thanks[X]{E-mail:partha.roychowdhury@saha.ac.in}
\thanks[Z]{E-mail:dnb@veccal.ernet.in}
\thanks[Y]{E-mail:chhanda.samanta@saha.ac.in}

\vskip -0.4cm
\begin{abstract}
\vskip -0.4cm

      The density dependence of nuclear symmetry energy is determined from a systematic study of the isospin dependent bulk properties of asymmetric nuclear matter using the isoscalar and the isovector components of density dependent M3Y interaction. The incompressibility $K_\infty$ for the symmetric nuclear matter, the isospin dependent part $K_{asy}$ of the isobaric incompressibility and the slope $L$ are all in excellent agreement with the constraints recently extracted from measured isotopic dependence of the giant monopole resonances in even-A Sn isotopes, from the neutron skin thickness of nuclei and from analyses of experimental data on isospin diffusion and isotopic scaling in intermediate energy heavy-ion collisions. This work provides a fundamental basis for the understanding of nuclear matter under extreme conditions, and validates the important empirical constraints obtained from recent experimental data.

\end{abstract}
\vskip -0.2cm
\noindent
\vskip -0.7cm
\end{frontmatter}

\section{Introduction}

      Measurements of nuclear masses, densities and collective excitations have allowed to resolve some of the basic features of the equation of state (EoS) of nuclear matter. However, the symmetry properties of the EoS due to differing neutron and proton numbers remain more elusive to date and study of the isospin dependent properties of asymmetric nuclear matter and the density dependence of the nuclear symmetry energy (NSE) remain the prime objective \cite{Li08,La07,St05,Ba05,Pi05,Pi07,Sh07,Fa06,Ch05,Br01}. Many radioactive ion beam facilities and their upgrades already exist and many more are being constructed or planned, including the Facility for Rare Isotope Beams (FRIB) in U.S.A. \cite{usa}, SPIRAL2/GANIL in France \cite{ganil}, FAIR/GSI in Germany \cite{gsi}, Radioactive Ion Beam (RIB) Factory at RIKEN in Japan \cite{Ya07} and the Cooling Storage Ring (CSR) facility at HIRFL in China \cite{Zh06}. These new facilities provide the possibility of exploring the properties of nuclear matter and nuclei under the extreme condition of large isospin asymmetry. Consequently, the study of the isospin degree of freedom in nuclear physics has recently received a lot of importance and attention. The ultimate goal of such study is to extract information on the isospin dependence of in-medium nuclear effective interactions as well as the EoS of isospin asymmetric nuclear matter, particularly its isospin-dependent term or the density dependence of the nuclear symmetry energy. This knowledge, especially the latter, is important for understanding not only the structure of radioactive nuclei, the reaction dynamics induced by rare isotopes, and the liquid-gas phase transition in asymmetric nuclear matter, but also many critical issues in astrophysics \cite{St05,Ba05,Li98,Da02}. With recent developments in constraining the isospin dependent properties of asymmetric nuclear matter, especially the density dependence of the nuclear symmetry energy, it is of great interest to see to what extent the results from a theoretical model are consistent with these constraints.

      In this work, based on the theoretical description of nuclear matter using the density dependent M3Y-Reid-Elliott effective interaction \cite{Be77,Sa79} (DDM3Y), we carry out a systematic study of the isospin-dependent bulk properties of asymmetric nuclear matter. In particular, we study the nuclear incompressibility $K_\infty$ for symmetric nuclear matter (SNM), the density dependence of the nuclear symmetry energy and extract the slope $L$ and the curvature $K_{sym}$ parameters of the nuclear symmetry energy and the isospin dependent part $K_{asy}$ of the isobaric incompressibility. We compare the results with the constraints recently extracted from analyses of the isospin diffusion data from heavy-ion collisions based on the isospin and momentum-dependent IBUU04 transport model with in-medium nucleon-nucleon (NN) cross sections \cite{Ch05,Ts04,Li05}, isoscaling analyses of isotope ratios in intermediate energy heavy-ion collisions \cite{SY07} and measured isotopic dependence of the giant monopole resonances (GMR) in even-A Sn isotopes \cite{Li07} and from the neutron skin thickness of nuclei \cite{Ce09}. 

\section{The nuclear equation of state and symmetry energy}

      The nuclear matter EoS is calculated using the isoscalar and the isovector components \cite{Sa83} of M3Y interaction along with density dependence. The density dependence of the effective interaction, DDM3Y, is completely determined from nuclear matter calculations. The equilibrium density of the nuclear matter is determined by minimizing the energy per nucleon. The energy variation of the zero range potential is treated accurately by allowing it to vary freely with the kinetic energy part $\epsilon^{kin}$ of the energy per nucleon $\epsilon$ over the entire range of $\epsilon$ \cite{BCS08}. This is not only more plausible, but also yields excellent result for the incompressibility $K_\infty$ of the SNM which does not suffer from the superluminosity problem.

      Assuming interacting Fermi gas of neutrons and protons, with isospin asymmetry $X= \frac{\rho_n - \rho_p} {\rho_n + \rho_p},~~~~\rho = \rho_n+\rho_p,$ where $\rho_n$, $\rho_p$ and $\rho$ are the neutron, proton and nucleonic densities respectively, the energy per nucleon for isospin asymmetric nuclear matter can be derived as \cite{BCS08}

\begin{equation}
 \epsilon(\rho,X) = [\frac{3\hbar^2k_F^2}{10m}] F(X) + (\frac{\rho J_v C}{2}) (1 - \beta\rho^n) 
\label{seqn1}
\end{equation}
\noindent
where $k_F= (1.5\pi^2\rho)^{\frac{1}{3}}$ which is equal to Fermi momentum in case of SNM, the kinetic energy per nucleon $\epsilon^{kin} = [\frac{3\hbar^2k_F^2}{10m}] F(X)$ with $F(X) = [\frac{(1+X)^{5/3} + (1-X)^{5/3}}{2}]$ and $J_v=J_{v00} + X^2 J_{v01}$ where $J_{v00}$ and $J_{v01}$ represent the volume integrals of the isoscalar and the isovector parts of the M3Y interaction supplemented by the zero-range potentials having the form    
 
\begin{eqnarray}
 && J_{v00} = J_{v00}(\epsilon^{kin})=\int \int \int t_{00}^{M3Y}(s, \epsilon) d^3s \nonumber \\ =&& 7999\frac{4\pi}{4^3} - 2134\frac{4\pi}{2.5^3} + J_{00} (1 - \alpha\epsilon^{kin})
 ~{\rm where}~J_{00}=-276~{\rm MeV~fm^3},
\label{seqn2}
\end{eqnarray}
\noindent

\begin{eqnarray}
 &&J_{v01} = J_{v01}(\epsilon^{kin}) = \int \int \int t_{01}^{M3Y}(s, \epsilon) d^3s \nonumber \\
=&& -4886\frac{4\pi}{4^3} + 1176\frac{4\pi}{2.5^3} + J_{01} (1 - \alpha\epsilon^{kin}) 
~{\rm where}~J_{01}=228~{\rm MeV~fm^3}.
\label{seqn3}
\end{eqnarray}
\noindent 

      The isoscalar $t_{00}^{M3Y}$ and the isovector $t_{01}^{M3Y}$ components of M3Y interaction potential \cite{Sa79,BCS08} supplemented by zero range potentials are given by $t_{00}^{M3Y}(s, \epsilon) = 7999\frac{\exp( - 4s)}{4s} - 2134\frac{\exp( - 2.5s)}{2.5s} - 276 (1 - \alpha\epsilon)\delta(s)$ and $t_{01}^{M3Y}(s, \epsilon) = -4886\frac{\exp( - 4s)}{4s} + 1176\frac{\exp( - 2.5s)}{2.5s} + 228 (1 - \alpha\epsilon)\delta(s)$ respectively, where the energy dependence parameter $\alpha$=0.005/MeV. The DDM3Y effective NN interaction is given by $v_{0i}(s,\rho, \epsilon) = t_{0i}^{M3Y}(s, \epsilon) g(\rho)$ where the density dependence $g(\rho) = C (1 - \beta \rho^n)$ and the constants $C$ and $\beta$ of density dependence have been obtained from the saturation condition $\frac{\partial\epsilon}{\partial\rho} = 0$ at $\rho = \rho_{0}$ and $\epsilon = \epsilon_{0}$ where $\rho_{0}$ and $\epsilon_{0}$ are the saturation density and the saturation energy per nucleon respectively. The equilibrium density of the cold SNM is determined from the saturation condition. Then Eq.(1) along with the equation for the saturation condition mentioned above can be solved simultaneously for fixed values of the saturation energy per nucleon $\epsilon_0$ and the saturation density $\rho_{0}$ of the cold SNM to obtain the values of $\beta$ and $C$. The constants of density dependence $\beta$ and $C$, thus obtained, are given by $\beta=\frac{[(1-p)+(q-\frac{3q}{p})]\rho_{0}^{-n}}{[(3n+1)-(n+1)p + (q-\frac{3q}{p})]}$ {\rm where} $p = \frac{[10m\epsilon_0]}{[\hbar^2k_{F_0}^2]}$, $q=\frac{2\alpha\epsilon_0J_{00}}{J^0_{v00}}$ with $J^0_{v00} = J_{v00}(\epsilon^{kin}_0)$ which means $J_{v00}$ at $\epsilon^{kin}=\epsilon^{kin}_0$, the kinetic energy part of the saturation energy per nucleon of SNM,  $k_{F_0} = [1.5\pi^2\rho_0]^{1/3}$ and $C = -\frac{[2\hbar^2k_{F_0}^2] }{ 5mJ^0_{v00} \rho_0 [1 - (n+1)\beta\rho_0^n -\frac{q\hbar^2k_{F_0}^2 (1-\beta\rho_0^n)} {10m\epsilon_0}]}$ respectively. It is quite obvious that the constants of density dependence $C$ and $\beta$ obtained by this method depend on the saturation energy per nucleon $\epsilon_0$, the saturation density $\rho_{0}$, the index $n$ of the density dependent part and on the strengths of the M3Y interaction through the volume integral $J^0_{v00}$. 

      The incompressibility or the compression modulus of the SNM, which is a measure of the curvature of an EoS at saturation density and defined as $k_F^2\frac{\partial^2\epsilon}{\partial{k_F^2}} \mid_{k_F=k_{F_0}}$, measures the stiffness of an EoS can be obtained theoretically:

\begin{eqnarray}
 K_\infty &&= k_F^2\frac{\partial^2\epsilon}{\partial{k_F^2}} \mid_{k_F=k_{F_0}} = 9\rho^2\frac{\partial^2\epsilon}{\partial\rho^2} \mid_{\rho=\rho_0} = -\frac{3\hbar^2k_{F_0}^2}{5m} - \frac{9 J^0_{v00} C n(n+1) \beta\rho_0^{n+1}}{2} \nonumber \\
&&- 9\alpha J_{00} C [1-(n+1)\beta\rho_0^n] \frac{\rho_0\hbar^2k_{F_0}^2}{5m}
+ \frac{3\rho_0\alpha J_{00} C (1-\beta\rho_0^n)\hbar^2k_{F_0}^2}{10m}.
\label{seqn4}
\end{eqnarray} 
\noindent

      The EoS of isospin asymmetric nuclear matter, given by Eq.(1) can be generally expanded as

\begin{equation}
  \epsilon(\rho,X) =  \epsilon(\rho,0) + E_{sym}(\rho) X^2 + O ( X^4)
\label{seqn5}
\end{equation} 
\noindent
and $E_{sym}(\rho)= \frac{1}{2} \frac{\partial^2\epsilon(\rho,X)}{\partial{X^2}} \mid_{X=0}$ is the nuclear symmetry energy. The absence of odd-order terms in $X$ in Eq.(5) is due to the exchange symmetry between protons and neutrons in nuclear matter when one neglects the Coulomb interaction and assumes the charge symmetry of nuclear forces. The higher-order terms in $X$ are negligible and as a good approximation, the density-dependent symmetry energy $E_{sym}(\rho)$ can be extracted from \cite{Kl06}

\begin{equation}
 E_{sym}(\rho)=\epsilon(\rho,1) -\epsilon(\rho,0)
\label{seqn6}
\end{equation}
\noindent
which can be obtained using Eq.(1) and represents a penalty levied on the system as it departs from the symmetric limit of equal number of protons and neutrons and can be defined as the energy required per nucleon to change the SNM to pure neutron matter (PNM).

\section{Isospin dependent properties of asymmetric nuclear matter}

      Around the nuclear matter saturation density $\rho_0$ the nuclear symmetry energy $E_{sym}(\rho)$ can be expanded to second order in density as 

\begin{equation}
 E_{sym}(\rho)= E_{sym}(\rho_0) + \frac{L}{3} {\Big (}\frac{\rho - \rho_0}{\rho_0}{\Big )}+ \frac{K_{sym}}{18}{\Big (}\frac{\rho - \rho_0}{\rho_0}{\Big )}^2
\label{seqn7}
\end{equation}
\noindent
where $L$ and $K_{sym}$ are the slope and curvature parameters of nuclear symmetry energy at $\rho_0$ and hence

\begin{equation}
 L= 3\rho_0 \frac{\partial E_{sym}(\rho)}{\partial\rho} \mid_{\rho=\rho_0}
\label{seqn8}
\end{equation}
\noindent

\begin{equation}
K_{sym}= 9\rho_0^2 \frac{\partial^2 E_{sym}(\rho)}{\partial {\rho^2}} \mid_{\rho=\rho_0}
\label{seqn9}
\end{equation}
\noindent
The $L$ and $K_{sym}$ characterize the density dependence of the nuclear symmetry energy around normal nuclear matter density and thus carry important information on the properties of nuclear symmetry energy at both high and low densities. In particular, the slope parameter $L$ has been found to correlate
linearly with the neutron-skin thickness of heavy nuclei and thus can in principle be determined from the measured thickness of neutron skin of such nuclei \cite{Pi05,Ch05,Br00,Ty01,Fu02,Ka02,Di03,SL05}. Recently, this has been possible \cite{Ce09} although there are large uncertainties in the experimental measurements.

      The Eq.(6) can be differentiated twice with respect to nucleonic density $\rho$ using Eq.(1) to yield  

\begin{eqnarray}
\vspace{0.0cm}
 \frac{\partial E_{sym}}{\partial \rho} &&=\frac{2}{5}(2^{2/3}-1)\frac{E^0_F}{\rho}(\frac{\rho}{\rho_0})^{2/3}+\frac{C}{2}[1-(n+1)\beta\rho^n] J_{v01}(\epsilon^{kin}_{X=1}) \nonumber \\
&&-\frac{\alpha J_{01}C}{5}E^0_F(\frac{\rho}{\rho_0})^{2/3}[1-\beta\rho^n]F(1) \nonumber \\
&&-(2^{2/3}-1)\frac{\alpha J_{00}C}{5}E^0_F(\frac{\rho}{\rho_0})^{2/3}[1-\beta\rho^n] \nonumber \\
&&-\frac{3}{10}(2^{2/3}-1)\alpha J_{00}CE^0_F(\frac{\rho}{\rho_0})^{2/3}[1-(n+1)\beta\rho^n]
\label{seqn10}
\end{eqnarray}
\vspace{0.7cm} 

\begin{eqnarray}
 \frac{\partial^2 E_{sym}}{\partial \rho^2} &&=-\frac{2}{15}(2^{2/3}-1)\frac{E^0_F}{\rho^2}(\frac{\rho}{\rho_0})^{2/3}-\frac{C}{2}n(n+1)\beta\rho^{n-1}J_{v01} (\epsilon^{kin}_{X=1}) \nonumber \\
&&-\frac{2\alpha J_{01}C}{5}\frac{E^0_F}{\rho}(\frac{\rho}{\rho_0})^{2/3}[1-(n+1)\beta\rho^n]F(1)  \nonumber \\
&&+\frac{\alpha J_{01}C}{15}\frac{E^0_F}{\rho}(\frac{\rho}{\rho_0})^{2/3}[1-\beta\rho^n]F(1)  \nonumber \\
&&+(2^{2/3}-1)\frac{\alpha J_{00}C}{15}\frac{E^0_F}{\rho}(\frac{\rho}{\rho_0})^{2/3} [1-\beta\rho^n] \nonumber \\
&&-\frac{2}{5}(2^{2/3}-1)\alpha J_{00}C\frac{E^0_F}{\rho}(\frac{\rho}{\rho_0})^{2/3} [1-(n+1) \beta\rho^n] \nonumber \\
&&+\frac{3}{10}(2^{2/3}-1)\alpha J_{00}CE^0_F(\frac{\rho}{\rho_0})^{2/3}n(n+1)\beta\rho^{n-1}
\label{seqn11}
\end{eqnarray} 
\noindent
where the Fermi energy $E^0_F=\frac{\hbar^2k_{F_0}^2}{2m}$ for the SNM at ground state. The above Eqs.(10,11) at $\rho$=$\rho_0$ are used to evaluate the values of $L$ and $K_{sym}$ using Eqs.(8,9). At the nuclear matter saturation density and around $X=0$, the isobaric incompressibility of asymmetric nuclear matter can also be expressed to second order in $X$ as \cite{Pr85,Lo88} $K(X) \approx K_\infty + K_{asy} X^2$ where $K_\infty$ is the incompressibility of symmetric nuclear matter at the nuclear matter saturation density, and the isospin dependent part \cite{Ba02}

\begin{equation}
K_{asy} \approx K_{sym}-6L
\label{seqn12}
\end{equation}
\noindent
characterizes the density dependence of the nuclear symmetry energy. Information on $K_{asy}$ can in principle be extracted experimentally by measuring the GMR in neutron-rich nuclei.

\section{Results of present calculations and experimental status}

      The calculations are performed using the values of the saturation density $\rho_0$=0.1533 fm$^{-3}$ \cite{Sa89} and the saturation energy per nucleon $\epsilon_0$=-15.26 MeV \cite{CB06} for the SNM obtained from the co-efficient of the volume term of Bethe-Weizs\"acker mass formula \cite{Bw35,Bw36} which is evaluated by fitting the recent experimental and estimated atomic mass excesses from Audi-Wapstra-Thibault atomic mass table \cite{Au03} by minimizing the mean square deviation incorporating correction for the electronic binding energy \cite{Lu03}. In a similar recent work, including surface symmetry energy term, Wigner term, shell correction and proton form factor correction to Coulomb energy also, $a_v$ turns out to be 15.4496 MeV \cite{Ro06} ($a_v$ =14.8497 MeV when $A^0$ and $A^{1/3}$ terms are also included). Using the usual values of $\alpha$=0.005 MeV$^{-1}$ for the parameter of energy dependence of the zero range potential and $n$=2/3, the values obtained for the constants of density dependence $C$ and $\beta$ and the SNM incompressibility $K_\infty$ are 2.2497, 1.5934 fm$^2$ and 274.7 MeV respectively. The saturation energy per nucleon is the volume energy coefficient and the value of -15.26$\pm$0.52 MeV covers, more or less, the entire range of values obtained for $a_v$ for which now the values of $C$=2.2497$\pm$0.0420, $\beta$=1.5934$\pm$0.0085 fm$^2$, the SNM incompressibility $K_\infty$=274.7$\pm$7.4 MeV, nuclear symmetry energy at nuclear density $E_{sym}(\rho_0)$=30.71 $\pm$0.26 MeV, the slope $L$=45.11$\pm$0.02 MeV  and the curvature $K_{sym}$=-183.7$\pm$3.6 MeV parameters of the nuclear symmetry energy and the isospin dependent part $K_{asy}$=-454.4$\pm$3.5 MeV of the isobaric incompressibility. These values are compared with those obtained from few other theoretical models in Table 1. In Fig.1 the NSE is plotted as a function of $\rho/\rho_0$ for the present calculation using DDM3Y interaction, and compared with those for Akmal-Pandharipande-Ravenhall \cite{Ak98} and MDI interaction \cite{Zh09}. 

\begin{figure}[htbp]
\vspace{2.5cm}
\eject\centerline{\epsfig{file=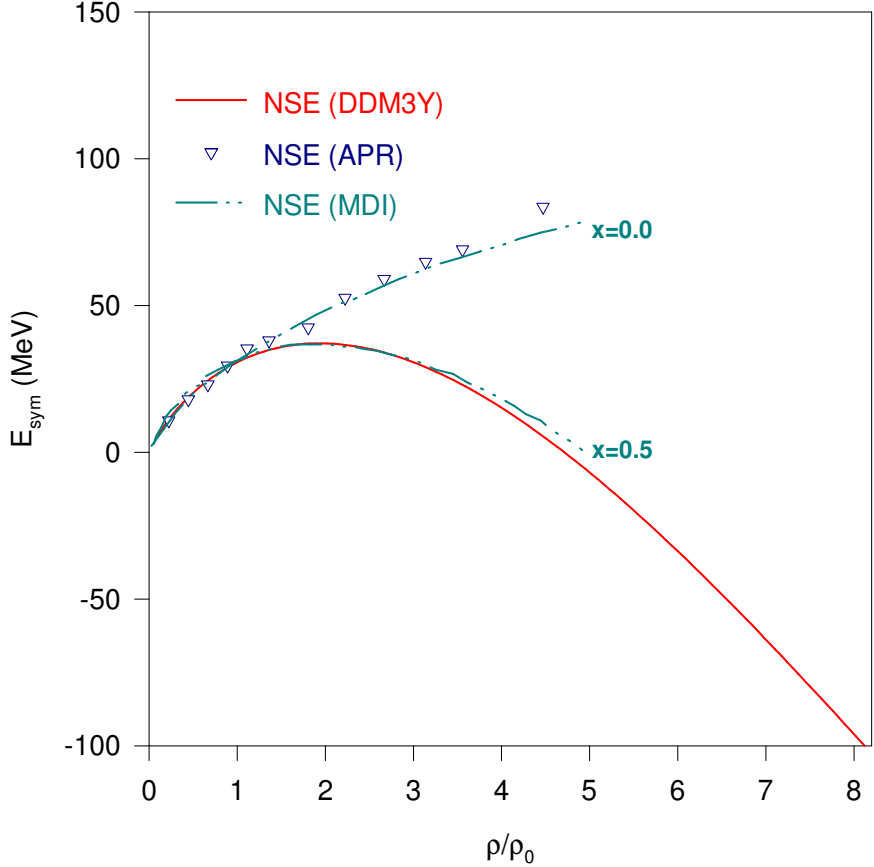,height=11.5cm,width=11.5cm}}
\caption
{The NSE (nuclear symmetry energy $E_{sym}$) is plotted as a function of $\rho/\rho_0$ for the present calculation using DDM3Y interaction and its comparison, with those for Akmal-Pandharipande-Ravenhall (APR) \cite{Ak98} and MDI interaction for the variable x=0.0, 0.5 defined in Ref. \cite{Zh09}.}
\label{fig1}
\vspace{0.0cm}
\end{figure}
 
      The theoretical estimate of $K_\infty$ from the refractive $\alpha$-nucleus scattering is about 240-270 MeV \cite{Kh97,Kh07}. The recent experimental determination of $K_\infty$ based upon the production of hard photons in heavy ion collisions which led to the experimental estimate of $K_\infty=290\pm50$ MeV \cite{Sc96}. However, the experimental values of $K_\infty$ extracted from the isoscalar giant dipole resonance (ISGDR) are claimed to be smaller \cite{Ga04}. Considering the status of experimental determination of the SNM incompressibility from data on the compression modes ISGMR and ISGDR of nuclei it can be inferred \cite{Sh06} that due to violations of self consistency in HF-RPA calculations of the strength functions of giant resonances result in shifts in the calculated values of the centroid energies which may be larger in magnitude than the current experimental uncertainties. In fact, the prediction of $K_\infty$ lying in the range of 210-220 MeV were due to the use of a not fully self-consistent Skyrme calculations \cite{Sh06}. Correcting for this drawback, Skyrme parmetrizations of SLy4 type predict $K_\infty$ values in the range of 230-240 MeV \cite{Sh06}. Moreover, it is possible to build bona fide Skyrme forces so that the SNM incompressibility is close to the relativistic value, namely 250-270 MeV. Therefore, from the ISGMR experimental data the conclusion can be drawn that $K_\infty \approx$ 240 $\pm$ 20 MeV. The ISGDR data tend to point to lower values \cite{Ga04,Lu04,Yo04} for $K_\infty$. However, there is consensus that the extraction of $K_\infty$ is in this case more problematic for various reasons. In particular, the maximum cross-section for ISGDR decreases very strongly at high excitation energy and may drop below the current experimental sensitivity for excitation energies \cite{Sh06} above 30 and 26 MeV for $^{116}$Sn and $^{208}$Pb, respectively. The present non-relativistic mean field model estimate for the nuclear incompressibility $K_\infty$ for SNM using DDM3Y interaction is rather close to the theoretical estimates obtained using relativistic mean field models and close to the upper limit of the recent values \cite{Yo05} extracted from experiments. The present consensus is that the acceptable value \cite{Vr03,Sh09} for the incompressibility of symmetric nuclear matter lies in the range of 250-270 MeV. The present value of 274.7$\pm$7.4 MeV for the incompressibility $K_\infty$ of SNM obtained using DDM3Y interaction is, therefore, an excellent theoretical result.  

      The constant of density dependence $\beta$=1.5934$\pm$0.0085 fm$^2$, which has the dimension of cross section for $n$=2/3, can be interpreted as the isospin averaged effective nucleon-nucleon interaction cross section in ground state symmetric nuclear medium. For a nucleon in ground state nuclear matter $k_F\approx$ 1.3 fm$^{-1}$ and $q_0 \sim \hbar k_F c \approx$ 260 MeV and the present result for the `in medium' effective cross section is reasonably close to the value obtained from a rigorous Dirac-Brueckner-Hartree-Fock \cite{Sa06} calculations corresponding to such $k_F$ and $q_0$ values which is $\approx$ 12 mb. Using the value of the constant of density dependence $\beta$=1.5934$\pm$0.0085 fm$^2$ corresponding to the standard value of the parameter $n$=2/3 along with the nucleonic density of 0.1533 fm$^{-3}$, the value obtained for the nuclear mean free path $\lambda$ is about 4 fm which is in excellent agreement \cite{Si83} with that obtained using another method.      

      The volume symmetry energy coefficient $S_v$ extracted from the masses of finite nuclei provides a constraint on the nuclear symmetry energy at nuclear density $E_{sym}(\rho_0)$. The value of $S_v=30.048 \pm 0.004$ MeV recently extracted \cite{Mu06} from the measured atomic mass excesses of 2228 nuclei is reasonably close to the theoretical estimate of the value of NSE at the saturation density $E_{sym}(\rho_0)$=30.71$\pm$0.26 MeV obtained from the present calculations using DDM3Y interaction. In ref. \cite{Da03} it is between 29.10 MeV to 32.67 MeV and that obtained by the liquid droplet model calculation of ref. \cite{St05} is 27.3 MeV whereas in ref. \cite{Di05} it is 28.0 MeV. It should be mentioned that the value of the volume symmetry parameter $S_v$ in some advanced mass description \cite{Po03} is close to the present value which with their $-\kappa_{vol}.b_{vol}=S_v$ equals 29.3 MeV. The value of NSE at nuclear saturation density $\approx$ 30 MeV, therefore, seems well established empirically. Theoretically different parameterizations of the relativistic mean-field (RMF) models, which fit observables for isospin symmetric nuclei well, lead to a relatively wide range of predictions 24-40 MeV for $E_{sym}(\rho_0)$. The present result of 30.71$\pm$0.26 MeV of the mean field calculation is close to the results of the calculation using Skyrme interaction SkMP (29.9 MeV) \cite{Be89}, Av18+$\delta v$+UIX$^*$ variational calculation (30.1 MeV) \cite{Ak98} and field theoretical calculation DD-F (31.6 MeV) \cite{Kl06}.   

\noindent 
\begin{table}[h]
\centering
\caption{Results of the present calculations (DDM3Y) of incompressibility of isospin symmetric nuclear matter $K_\infty$, nuclear symmetry energy at saturation density $E_{sym}(\rho_0)$, the slope $L$ and the curvature $K_{sym}$ parameters of the nuclear symmetry energy and the isospin dependent part $K_{asy}$ of the isobaric incompressibility (all in MeV) are compared with those obtained with other models.}
\begin{tabular}{ccccccc}
\hline
\hline
Model&$K_\infty$&$E_{sym}(\rho_0)$&$L$&$K_{sym}$&$K_{asy}$&Ref. \\ 
\hline
 Expt. &$--$&$31.6$&$75\pm 25$&$--$&$-500^{+125}_{-100}$& \cite{Ce09} \\ 
 This work &$274.7\pm7.4$&$30.71\pm0.26$&$45.11\pm0.02$&$-183.7\pm3.6$&$-454.4\pm3.5$& \\ 
 NL1&212&43.5&140&143&-697&\cite{Le86} \\
 NL2&401&44.0&130&20&-750&\cite{Le86} \\
 NL3&271&37.3&118&100&-608&\cite{La97} \\
 NL-SH&356&36.1&114&80&-604&\cite{Sh93} \\
 FSUGold&229&32.5&60&-52&-412&\cite{Pi05} \\
 DD-ME1&245&33.1&55&-101&-431&\cite{Ni02} \\
 DD-ME2&251&32.3&51&-87&-393&\cite{La05} \\
 DD &241&31.7&56&-95&-431&\cite{Ty05} \\
 DD-F&223&31.6&56&-140&-476&\cite{Kl06} \\
  \hline
\hline
\end{tabular} 
\end{table}

      Information on $K_{asy}$ can in principle be extracted experimentally by measuring the GMR in neutron-rich nuclei. Earlier attempts based on this method have given, however, widely different values. A value of $K_{asy}=-320\pm$180 MeV \cite{Sh88} with a large uncertainty was obtained from a systematic study of the GMR in the isotopic chains of Sn and Sm. In this analysis, the value of $K_\infty$ was found to be 300$\pm$25 MeV. The recently obtained values from the isoscaling analyses of isotope ratios in intermediate energy heavy-ion collisions \cite{SY07} gives $L\approx$65 MeV and $K_{asy}\approx-453$ MeV. The extracted values of $L=88\pm$25 MeV and $K_{asy}=-500\pm$50 MeV from the isospin diffusion data \cite{Ch07} and $K_{asy}=-550\pm$100 MeV obtained from recently measured isotopic dependence of the GMR in even-A Sn isotopes \cite{Li07} are consistent with the value of $K_{asy}$ obtained from the present calculations. The incompressibility of symmetric nuclear matter at its saturation density has been determined to be 240$\pm$20 MeV \cite{Ch09} from analyses of the GMR. A very recent extraction from neutron skin thickness of nuclei provides the values of $L$=75$\pm$25 MeV (revised to narrower window of L$\sim$45-75 MeV recently \cite{Wa09}) and $K_{asy}$=$-500^{+125}_{-100}$ MeV \cite{Ce09} and it is needless to stress that the theoretical values of our calculations for $L$=45.11$\pm$0.02 MeV and $K_{asy}$=$-454.4\pm3.5$ MeV are in excellent agreement.

\section{Summary and conclusion}

      We show that the theoretical description of nuclear matter based on mean field calculation using DDM3Y effective NN interaction yields a value of nuclear incompressibility which is highly in agreement with that extracted from experiment and gives a value of symmetry energy that is consistent with the empirical value extracted by fitting the droplet model to the measured atomic mass excesses and with other modern theoretical descriptions of nuclear matter. The present NSE is `soft' because it increases initially with nucleonic density up to about two times the normal nuclear density and then decreases monotonically at higher densities and is consistent with the recent evidence for a soft NSE at suprasaturation densities \cite{Zh09}. This interaction also provides good descriptions for the elastic and inelastic scattering \cite{Gu06} and various kinds of radioactivities \cite{Ba07}. The slope $L$ and the isospin dependent part $K_{asy}$ of the isobaric incompressibility are consistent with the constraints recently extracted from analyses of experimental data. Of all other models, DD-ME2 provides comparatively better estimates and of all RMF models provides best mass predictions with r.m.s. deviation of 0.9 MeV. Interestingly, our calculations provide so far the best theoretical estimates for the isospin dependent properties of asymmetric nuclear matter.


\begin{thebibliography}{999}

\bibitem{Li08} B. A. Li, L. W. Chen, and C. M. Ko, Phys. Rep. {\bf 464} (2008) 113.

\bibitem{La07} J. M. Lattimer and M. Prakash, Phys. Rep. {\bf 442} (2007) 109.

\bibitem{St05} A. W. Steiner, M. Prakash, J. M. Lattimer and P. J. Ellis, Phys. Rep. {\bf 411} (2005) 325. 

\bibitem{Ba05} V. Baran, M. Colonna, V. Greco, and M. Di Toro, Phys. Rep. {\bf 410} (2005) 335.

\bibitem{Pi05} C. J. Horowitz and J. Piekarewicz, Phys. Rev. Lett. {\bf 86} (2001) 5647; B. G. Todd-Rutel and J. Piekarewicz, Phys. Rev. Lett. {\bf 95} (2005) 122501.

\bibitem{Pi07} J. Piekarewicz, Phys. Rev. {\bf C 76} (2007) 031301(R).

\bibitem{Sh07} D.V. Shetty, S. J. Yennello, and G. A. Souliotis, Phys. Rev. {\bf C 76} (2007) 024606.

\bibitem{Fa06} M. A. Famiano, T. Liu, W. G. Lynch, M. Mocko, A. M. Rogers, M. B. Tsang, M. S. Wallace, R. J. Charity, S. Komarov, D. G. Sarantites, L. G. Sobotka, and G. Verde, Phys. Rev. Lett. {\bf 97} (2006) 052701.

\bibitem{Ch05} L. W. Chen, C. M. Ko, and B. A. Li, Phys. Rev. Lett. {\bf 94} (2005) 032701; Phys. Rev. {\bf C 72} (2005) 064309.

\bibitem{Br01} B. A. Brown, Phys. Rev. Lett. {\bf 85} (2000) 5296; S. Typel and B. A. Brown, Phys. Rev. {\bf C 64}  (2001) 027302.

\bibitem{usa} White Papers of the 2007 NSAC Long Range Plan Town Meeting, Jan. 2007, Chicago, http://dnp.aps.org.

\bibitem{ganil} http://ganinfo.in2p3.fr/research/developments/spiral2.

\bibitem{gsi} http://www.gsi.de/fair/index e.html.

\bibitem{Ya07} Y. Yano, Nucl. Instrum. Methods {\bf B 261} (2007) 1009.

\bibitem{Zh06} W. Zhan et al., Int. J. Mod. Phys. {\bf E 15} (2006) 1941; http://www.impcas.ac.cn/zhuye/en/htm/247.htm.

\bibitem{Li98} B. A. Li, C. M. Ko, and W. Bauer, Int. J. Mod. Phys. {\bf E 7} (1998) 147.

\bibitem{Da02} P. Danielewicz, R. Lacey and W.G. Lynch, Science {\bf 298} (2002) 1592. 

\bibitem{Be77} G.Bertsch, J.Borysowicz, H.McManus, W.G.Love, Nucl. Phys. {\bf A 284} (1977) 399.

\bibitem{Sa79} G.R. Satchler and W.G. Love, Phys. Reports {\bf 55} (1979) 183. 

\bibitem{Ts04} M. B. Tsang et al., Phys. Rev. Lett. {\bf 92} (2004) 062701.

\bibitem{Li05} B. A. Li and L. W. Chen, Phys. Rev. {\bf C 72} (2005) 064611.

\bibitem{SY07} D. V. Shetty, S. J. Yennello, and G. A. Souliotis, Phys. Rev. {\bf C 75} (2007) 034602.

\bibitem{Li07} T. Li, U. Garg, Y. Liu et al., Phys. Rev. Lett. {\bf 99} (2007) 162503.

\bibitem{Ce09} M. Centelles, X. Roca-Maza, X. Vinas and M. Warda, Phys. Rev. Lett. {\bf 102} (2009) 122502.

\bibitem{Sa83} A.M. Lane, Nucl. Phys. {\bf 35} (1962) 676; G.R. Satchler, {\it Int. series of monographs on Physics}, Oxford University Press, {\it Direct Nuclear reactions} (1983) 470.

\bibitem{BCS08} D. N. Basu, P. Roy Chowdhury and C. Samanta, Nucl. Phys. {\bf A 811} (2008) 140.

\bibitem{Kl06} T. Kl\"ahn et al., Phys. Rev. {\bf C 74} (2006) 035802. 

\bibitem{Br00} B. A. Brown, Phys. Rev. Lett. {\bf 85} (2000) 5296.

\bibitem{Ty01} S. Typel and B. A. Brown, Phys. Rev. {\bf C 64} (2001) 027302.

\bibitem{Fu02} R. J. Furnstahl, Nucl. Phys. {\bf A 706} (2002) 85.

\bibitem{Ka02} S. Karataglidis, K. Amos, B. A. Brown and P. K. Deb, Phys. Rev. {\bf C 65} (2002) 044306.

\bibitem{Di03} A. E. L. Dieperink, Y. Dewulf, D. Van Neck, M. Waroquier and V. Rodin, Phys. Rev. {\bf C 68} (2003) 064307.

\bibitem{SL05} A. W. Steiner and B. A. Li, Phys. Rev. {\bf C 72} (2005) 041601(R).

\bibitem{Pr85} M. Prakash and K. S. Bedell, Phys. Rev. {\bf C 32} (1985) 1118.

\bibitem{Lo88} M. Lopez-Quelle, S. Marcos, R. Niembro, A. Bouyssy and N. V. Giai, Nucl. Phys. {\bf A 483} (1988) 479.

\bibitem{Ba02} V. Baran, M. Colonna, M. Di Toro, V. Greco, and M. Zielinska-Pfab\'e and H. H. Wolter, Nucl. Phys. {\bf A 703} (2002) 603.

\bibitem{Sa89} C. Samanta, D. Bandyopadhyay and J.N. De, Phys. Lett. {\bf B 217} (1989) 381. 

\bibitem{CB06} P. Roy Chowdhury and D.N. Basu, Acta Phys. Pol. {\bf B 37} (2006) 1833.

\bibitem{Bw35} C.F. Weizs\"acker, Z.Physik {\bf 96} (1935) 431.

\bibitem{Bw36} H.A. Bethe, R.F. Bacher, Rev. Mod. Phys. {\bf 8} (1936) 82.

\bibitem{Au03} G. Audi, A.H. Wapstra and C. Thibault, Nucl. Phys. {\bf A 729} (2003) 337.

\bibitem {Lu03} D. Lunney, J.M. Pearson and C. Thibault, Rev. Mod. Phys. {\bf 75} (2003) 1021.

\bibitem{Ro06} G.Royer and C.Gautier, Phys. Rev. {\bf C 73} (2006) 067302.

\bibitem{Ak98} A. Akmal, V. R. Pandharipande and D. G. Ravenhall, Phys. Rev. {\bf C 58} (1998) 1804.

\bibitem{Zh09} Zhigang Xiao, Bao-An Li, Lie-Wen Chen, Gao-Chan Yong and Ming Zhang, Phys. Rev. Lett. {\bf 102}, 062502 (2009).

\bibitem{Kh97} Dao T. Khoa, G.R. Satchler and W. von Oertzen, Phys. Rev. {\bf C 56} (1997) 954. 

\bibitem{Kh07} Dao T. Khoa, W. von Oertzen, H.G. Bohlen and S. Ohkubo, Jour. Phys. {\bf G 34} (2007) R111.    

\bibitem{Sc96} Y. Schutz et al., Nucl. Phys. {\bf A 599} (1996) 97c.

\bibitem{Ga04} U. Garg, Nucl. Phys. {\bf A 731} (2004) 3. 

\bibitem{Sh06} S. Shlomo, V. M. Kolomietz and G. Colo, Eur. Phys. Jour. {\bf A 30} (2006) 23. 

\bibitem{Lu04} Y. W. Lui, D. H. Youngblood, Y. Tokimoto,  H. L. Clark and B. John, Phys. Rev. {\bf C 69} (2004) 034611.

\bibitem{Yo04} D. H. Youngblood, Y. W. Lui, B. John, Y. Tokimoto,  H. L. Clark and X. Chen, Phys. Rev. {\bf C 69} (2004) 054312.
 
\bibitem{Yo05} Y. W. Lui, D.H. Youngblood, H.L. Clark, Y. Tokimoto and B. John, Acta Phys. Pol. {\bf B 36} (2005) 1107.

\bibitem{Vr03} D. Vretenar, T. Nik\'si\'c and P. Ring, Phys. Rev. {\bf C 68} (2003) 024310.

\bibitem{Sh09} M. M. Sharma, Nucl. Phys. {\bf A 816} (2009) 65.

\bibitem{Sa06} F. Sammarruca and P. Krastev, Phys. Rev. {\bf C 73} (2006) 014001.

\bibitem{Si83} B. Sinha, Phys. Rev. Lett. {\bf 50} (1983) 91.

\bibitem{Mu06} T. Mukhopadhyay and D.N. Basu, Acta Phys. Pol. {\bf B 38} (2007) 3225.

\bibitem{Da03} P. Danielewicz, Nucl. Phys. {\bf A 727} (2003) 233. 

\bibitem{Di05} A.E.L. Dieperink and D van Neck, Jour. Phys. Conf. Series {\bf 20} (2005) 160.

\bibitem{Po03} K. Pomorski and J. Dudek, Phys. Rev. {\bf C 67} (2003) 044316. 

\bibitem{Be89} L. Bennour et al., Phys. Rev. C {\bf 40} (1989) 2834.

\bibitem{Le86} S.-J. Lee, J. Fink, A. B. Balantekin et al., Phys. Rev. Lett. {\bf 57} (1986) 2916.

\bibitem{La97} G. A. Lalazissis, J. Konig and P. Ring, Phys. Rev. {\bf C 55} (1997) 540.

\bibitem{Sh93} M. M. Sharma, M. A. Nagarajan and P. Ring, Phys. Lett. {\bf B 312} (1993) 377.

\bibitem{Ni02} T. Nik\'si\'c, D. Vretenar, P. Finelli and P. Ring, Phys. Rev. {\bf C 66} (2002) 024306.

\bibitem{La05} G. A. Lalazissis, T. Nik\'si\'c, D. Vretenar and P. Ring, Phys. Rev. {\bf C 71} (2005) 024312.

\bibitem{Ty05} S. Typel, Phys. Rev. {\bf C 71} (2005) 064301.

\bibitem{Sh88} M.M. Sharma, W. T. A. Borghols, S. Brandenburg, S. Crona, A. van der Woude and M. N. 
Harakeh, Phys. Rev. {\bf C 38} (1988) 2562. 

\bibitem{Ch07} Lie-Wen Chen, Che Ming Ko and Bao-An Li, Phys. Rev. {\bf C 76} (2007) 054316.

\bibitem{Ch09} Lie-Wen Chen, Bao-Jun Cai, Che Ming Ko, Bao-An Li, Chun Shen and Jun Xu, arXiv: 0905.4323.

\bibitem{Wa09} M. Warda, X. Vi\"nas, X. Roca-Maza and M. Centelles, arXiv: 0906.0932.

\bibitem{Gu06} D. Gupta and D. N. Basu, Nucl. Phys. {\bf A 748} (2005) 402; D. Gupta, E. Khan and  Y. Blumenfeld, Nucl. Phys. {\bf A 773} (2006) 230.  

\bibitem{Ba07} D. N. Basu, P. Roy Chowdhury and C. Samanta, Phys. Rev. {\bf C 72} (2005) 051601(R); C. Samanta, P. Roy Chowdhury and D. N. Basu, Nucl. Phys. {\bf A 789} (2007) 142; D. N. Basu, Phys. Lett. {\bf B 566} (2003) 90.

\end{thebibliography}
\end{document}